\documentclass[conference]{IEEEtran}
\usepackage[margin=25mm]{geometry}

\usepackage{amsfonts}
\usepackage{amssymb}

\usepackage{verbatim}
\usepackage{amsmath, amsthm}
\usepackage{graphicx}
\usepackage{float}

\providecommand{\keywords}[1]
{
  \small	
  \textbf{\textit{Keywords---}} #1
}

\newtheorem{dfn}{Definition}

\begin{document}

\title{PAMMELA: Policy Administration Methodology using Machine Learning}

\author{Varun Gumma$^{1}$, Barsha Mitra$^{2}$, Soumyadeep Dey$^{3}$,  Pratik Shashikantbhai Patel$^{2}$,  \\Sourabh Suman$^{2}$, Saptarshi Das$^{4}$\\
        \small $^{1}$Department of Computer Science and Engineering, IIT Madras, Chennai, India \\
        \small $^{2}$Department of CSIS, BITS Pilani, Hyderabad Campus, Hyderabad, India \\
        \small $^{3}$ Microsoft IDC, India \\
        \small $^{4}$ JIS Institute of Advanced Studies and Research, JIS University, Kolkata, India \\
}

\maketitle
\begin{abstract}
In recent years, Attribute-Based Access Control (ABAC) has become quite popular and effective for enforcing access control in dynamic and collaborative environments. Implementation of ABAC requires the creation of a set of attribute-based rules which cumulatively form a policy. Designing an ABAC policy ab initio demands a substantial amount of effort from the system administrator. Moreover, organizational changes may necessitate the inclusion of new rules in an already deployed policy. In such a case, re-mining the entire ABAC policy will require a considerable amount of time and administrative effort. Instead, it is better to incrementally augment the policy. Keeping these aspects of reducing administrative overhead in mind, in this paper, we propose PAMMELA, a Policy Administration Methodology using Machine Learning to help system administrators in creating new ABAC policies as well as augmenting existing ones. PAMMELA can generate a new policy for an organization by learning the rules of a policy currently enforced in a similar organization. For policy augmentation, PAMMELA can infer new rules based on the knowledge gathered from the existing rules. Experimental results show that our proposed approach provides a reasonably good performance in terms of the various machine learning evaluation metrics as well as execution time.
\end{abstract}


\section{Introduction}  \label{intro}
In any organization, it is of utmost importance to ensure that all accesses to resources take place in an authorized manner. This can be facilitated through an access control model. Over the years, several access control models have been proposed like the Discretionary Access Control (DAC) model \cite{dac}, the Mandatory Access Control (MAC) model \cite{mac} and the Role-Based Access Control (RBAC) \cite{rbac} model. Over the years, RBAC became quite popular as an effective means of access control. The model has been successfully deployed in various organizations and has been incorporated in a number of products and platforms. The pivotal element of RBAC is \textit{roles}. 

In spite of its widespread popularity, the RBAC model suffers from a major drawback of being unsuitable for dynamic environments. In such environments, it is not possible to know apriori the set of all users and the different types of access requests that may occur. Moreover, in situations where two or more organizations interact in a collaborative manner, RBAC cannot be implemented to enforce access control among the different organizations. The reason behind this is that the role definitions (in terms of the included permissions) of various organizations may not be the same. Such situations may require the creation of temporary roles in order to facilitate the collaborative interaction.

The Attribute-Based Access Control (ABAC) model \cite{abac_nist} was proposed for facilitating access control in dynamic and collaborative environments. ABAC encompasses the properties as well as the benefits of DAC, MAC and RBAC. This model allows users to access resources based on the properties of the users, resources and the environment. As per the terminologies of ABAC, the users, the resources and the properties are referred to as subjects, objects and attributes respectively. Each attribute is assigned a value or multiple values from a pre-defined set of values for every subject, object and environmental condition. In order to determine whether to allow or deny an access request, the attributes of the requesting subject, requested object as well as those of the environment in which the access request is made are taken into account. 
If each attribute has been assigned a specific value, then the access request is allowed, otherwise, it is denied. In order to deploy ABAC, a set of rules is required. A rule defines the required attributes and the corresponding permissible values for each attribute for a specific type of access. The set of all rules collectively comprise a \textit{policy}. 

The process of creating a policy for implementing ABAC is known as \textit{policy engineering}. Policy engineering can be carried out in two ways - \textit{top-down} \cite{topdown} and \textit{bottom-up} \cite{constrainedmining}, \cite{abac_policy_engg}, \cite{logdeep}, \cite{mining}. Bottom-up policy engineering is also termed as \textit{policy mining}. In the current literature, the policy mining problem has been formulated as a minimization problem. 
Combining top-down and bottom-up approaches gives \textit{hybrid policy engineering} \cite{hype}. 


An organization intending to migrate to ABAC requires designing an ABAC policy. Also, when an organization having an already deployed ABAC model, undergoes some changes (like the opening of a new department or the introduction of a new academic course), additional ABAC rules reflecting the changes need to be generated. Creating ABAC rules ab initio or completely re-mining an ABAC policy requires a considerable amount of administrative effort and time. These overheads can be substantially reduced if an existing policy is used as a reference for the creation of a new ABAC policy for migration. Also, instead of re-mining the complete policy in order to account for the organizational changes, it is more prudent to only create the additional rules. In both the scenarios, the existing ABAC policy can aid the process of creation of new ABAC rules, thereby reducing the effort and overhead of policy administration.

In this paper, we propose a methodology that will aid ABAC system administrators in efficiently augmenting an existing policy by including additional rules in order to accommodate various organizational alterations as well as generating a new ABAC policy for an organization by referring to the existing policy of a similar organization. Our proposed approach focuses on policy creation through a technique which is based on the information contained in an existing policy. The strategy makes use of machine learning techniques that are trained using the currently deployed policy. After training, our methodology generates new access rules from a set of access requests that they are not covered by the existing rules. 

The main contributions of the paper are summarized below.
\begin{itemize}
\item We propose the \textit{ABAC Policy Inference Problem} (ABAC-PIP) which takes as inputs an existing ABAC policy and a set of access requests and creates a new set of ABAC rules that either augments the existing policy or creates a new policy. It is to be noted here that  the rules that are created are considered as new since they are different from the existing ones in terms of certain attribute values.

\item We propose a machine learning based methodology for solving ABAC-PIP that makes use of supervised learning. We train a machine learning classifier using the existing ABAC policy. The training includes both positive (rules that grant accesses) as well as negative (rules that deny accesses) rules. After training, the classifier is supplied with a set of access requests (ones that should be allowed and ones that should be disallowed). Based on these, the machine learning algorithm generates a new set of access rules. 
We name our approach as \textit{\textbf{\underline{P}}olicy \textbf{\underline{A}}dministration \textbf{\underline{M}}ethodology using \textbf{\underline{M}}achin\textbf{\underline{e}} \textbf{\underline{L}}e\textbf{\underline{a}}rning} (PAMMELA).




\item We propose two techniques in order to enhance the performance of PAMMELA based the intrinsic relations that exist among the different attributes and the grouping of similar attribute values.

\item We test our proposed policy inferring methodology on three manually crafted datasets. These datasets have been carefully created by keeping in mind real-world scenarios. Our experiments show that PAMMELA shows promising results and provides a high degree of performance. We have experimented with a number of machine learning classifiers and show a comparative result for all them.
\end{itemize}

The rest of the paper is organized as follows. Section \ref{related} explores the different policy mining approaches present in the existing literature. In Section \ref{background}, we review the preliminary concepts related to the ABAC model and supervised learning. We formulate the problem definition of ABAC-PIP in Section \ref{problem} and present the corresponding solution strategy PAMMELA along with a discussion of the application scenarios and the learning enhancement techniques in Section \ref{framework}. Dataset description, evaluation metrics and the experimental results are presented in Section \ref{exp}. Finally, Section \ref{concl} concludes the paper with some insights into future research directions.

\section{Related Work} \label{related}

A considerable number of work has focused on developing techniques for creation of ABAC policies. Xu and Stoller proposed one of the earliest ABAC policy mining algorithms from access logs \cite{logs}. Another work by Xu and Stoller aims at generation of ABAC policies from access control lists and attribute data given as inputs \cite{mining}. They have formulated the ABAC policy mining problem as an optimization problem and have proposed \textit{weighted structural complexity} as the policy quality metric. Talukdar et al. \cite{efficient_mining} have proposed ABAC-SRM, a bottom-up policy mining method capable of creating generalized ABAC rules. Cotrini et al., in \cite{cotrini}, have proposed Rhapsody, a policy mining technique that can handle sparse inputs and have defined a rule quality metric called \textit{reliability}. 

An attribute-based rule mining algorithm has been proposed in \cite{abac_least_privilege}  from the audit logs of an organization that can minimize the under and over privileges for enforcing the principle of least privilege. The authors also propose a scoring method for determining the quality of a policy from a least privilege point of view. \cite{constraint_nlp} proposes a methodology that can extract ABAC constraints in an automated manner from policies expressed in natural language. A constrained policy mining technique has been proposed in \cite{constrainedmining}. In \cite{abac_policy_engg}, the authors have proposed a policy engineering approach that considers the risk associated with the improper use and possible abuse of a permitted access to a user. Lawal and Krishnan have proposed an approach for policy administration in ABAC via policy review \cite{policy_review}.  Heutelbeck et al. \cite{policy_indexing} have designed a data structure for efficiently indexing policy documents and have proposed a method for finding the relevant policy documents for a particular access request.

Several incremental and adaptive policy generation techniques are present in the current literature. Das et al. \cite{saptarshi_adaptation} have proposed a policy adaptation strategy between similar organizations. In this context, they have formulated the Policy Adaptation Problem (PolAP) which aims at determining the value assignments of the attributes of each subject for a given ABAC policy. They have proved the problem to be NP-complete and have proposed a heuristic algorithm for solving it. It can be noted that our proposed ABAC-PIP is different from PolAP both in terms of the inputs and the output. Das et al. have further extended their work in \cite{saptarshi_hierarchical} by considering hierarchical relationships among subject attribute values and also taking into account environmental attributes. Batra et al. \cite{incr_abac} have proposed an incremental policy mining technique that is capable of creating new ABAC rules in the event of any one of the following occurrences - (i) addition of a permission, (ii) deletion of a permission, (iii) addition of an attribute value and (iv) deletion of an attribute value. In \cite{policy_recon}, the authors have presented a strategy for determining policy similarity, have proposed two methods for performing policy reconciliation and also presented a policy migration technique. 

The rapid growth of artificial intelligence has ushered in a widespread application of different machine learning and deep learning algorithms in the field of access control. In \cite{logdeep}, the authors have presented a policy generation technique using Restricted Boltzmann Machines. Karimi et al. in \cite{joshi_log} have proposed an automated policy learning method from access logs using unsupervised learning by considering both positive and negative rules. Their approach can handle noisy data as well as sparseness of logs. The authors have also proposed rule pruning and policy refinement techniques. \cite{polisma} presents a framework known as \textit{Polisma} for learning ABAC policies from access logs by using a combination of statistical, data mining and machine learning algorithms. In \cite{abac_reinforcement}, the authors have designed a policy learning method that is adaptive in nature using a feedback loop and is applicable for home Internet of Things (IoT) environment. They have modeled the problem of ABAC policy learning as a reinforcement learning problem. Very recently, Bertino et al. have proposed an approach known as FLAP \cite{flap} for collaborative environments. FLAP enables one organization to learn policies from another organization and perform policy adaptation via a policy learning framework by using local log or local policies or local learning or hybrid learning.
This work makes use of Polisma \cite{polisma}.

In this work, we employ machine learning algorithms for the purpose of generating ABAC rules. We have also formulated the corresponding policy creation problem variant and have named it as the ABAC Policy Inference Problem (ABAC-PIP). To the best of our knowledge, such kind of problem formulation and the design of end-to-end machine learning based solution strategy have rarely been addressed in the existing literature.

\section{Background} \label{background}

In this section, we present some preliminaries related to the ABAC model and supervised learning techniques.

\subsection{ABAC Model} \label{abac}
The ABAC model consists of the following components:
\begin{itemize}
\item A set $S$ of subjects or users. Each subject 
can be a human being or a non-human entity.
\item A set $O$ of objects. Each object 
corresponds to a system resource that should be accessed in an authorized manner.
\item A set $E$ of environmental factors or conditions. Each condition 
can represent some temporal or spatial or some other kind of context in which a user requests access to a resource. Examples of environmental conditions can be location, time, temperature, etc.
\item A set $SA$ of subject attributes. Each subject attribute represents a property associated with a subject and can assume a single or multiple values from a set of values. These values are known as subject attribute values. If a subject attribute assumes a single value, it is known as \textit{atomic valued}. If it is assigned multiple values for a subject simultaneously, it is known as \textit{multi-valued}. An example of a subject attribute is $Department$. The attribute value set of $Department$ can include {\textit{Computer Science}, \textit{Electronics}, and \textit{Mechanical}}. Sometimes a specific subject attribute of a user may not be meaningful. Such a scenario can be represented by assigning the value \textit{Not Applicable} for that attribute.
\item A set $OA$ of object attributes. All the concepts mentioned for subject attributes are applicable for object attributes as well. An example of an object attribute is \textit{Type of Document} and its possible values can be \textit{Project Plan}, \textit{Budget}, \textit{Expenditure Details} etc.
\item A set $EA$ of environmental attributes. All the concepts discussed for subject attributes also apply for environmental attributes. Example of an environmental attribute can be \textit{Time of Doctor's Visit} and the possible values can be \textit{Day Shift} and \textit{Night Shift}.
\item A function $F_{sub}$ that assigns values to subject attributes of a subject. Formally, $F_{sub}$: $S$ x $SA$ $\rightarrow$ \{$v_s$ $|$ $v_s$ is a subject attribute value\}. For eg., $F_{sub}$($John$, $Department$) = \{$Computer\ Science$\}.
\item A function $F_{obj}$ that assigns values to object attributes of each object. Formally, $F_{obj}$: $O$ x $OA$ $\rightarrow$ \{$v_o$ $|$ $v_o$ is an object attribute value\}. For eg., $F_{obj}$($File1.doc$, $Type\ of \ Documet$) = \{$Project\ Plan$\}.
\item A function $F_{env}$ that assigns values to environmental attributes. Formally, $F_{env}$: $E$ x $EA$ $\rightarrow$ \{$v_e$ $|$ $v_e$ is an environmental attribute value\}. 
\item A set $P$ of operations or permissions. Common types of operations/permissions include \textit{read}, \textit{write}, \textit{update}, \textit{execute}, etc.
\item A set $\mathcal{R}$ of rules. Each rule specifies whether a particular type of access is to be granted or denied. A rule that permits an access is termed as a positive rule and a rule that disallows an access is termed as a negative rule. All the rules cumulatively constitute an ABAC policy. An example ABAC rule can be of the following form - <\{$Department$ = $Accounts$, $Designation$ = $Accountant$\}, \{$Type$ = $Payroll\ Data$, $Department$ = $Any$\}, $view$>. In natural language, this rule translates to - \textit{If a subject belongs to the Accounts department and has a Designation of Accountant, then she can view the payroll data of any employee belonging to any department}. Here, the value $Any$ is used to express the rule in a generalized form.
\end{itemize}
\subsection{Supervised Learning} \label{sl}
Supervised learning is a class of machine learning algorithms that are used either to classify data or to predict some type of outcomes based on labeled data. The machine learning model is trained using the labeled data. This training enables the model to learn the class type or the outcome for a given combination of parameter values. The trained model is used to assign the class labels or predict the outcomes of unlabeled data points. Supervised learning can address two types of problems - \\
\textbf{1.} \textbf{Classification}: Here the machine learning algorithm has to categorize inputs (or data points) into different classes. Eg., Support Vector Machine \cite{Murphy}, k-Nearest Neighbour \cite{Murphy}, Decision Tree \cite{DecisionTrees}, Random Forest \cite{RandomForest}, GradientBoost \cite{scikit-learn}, XGBoost \cite{scikit-learn}, etc.\\
\textbf{2.} \textbf{Regression}: Here the machine learning algorithm makes predictions for unknown data points. Examples include Linear Regression \cite{Murphy}, Logistic Regression \cite{Murphy}.

In this work, we have used only classification algorithms.

\section{ABAC Policy Inference Problem} \label{problem}

Designing ABAC policies is not a trivial task and requires a considerable amount of administrative effort. The overhead associated with the process of policy generation can be reduced if some existing policy serves as a reference point or guideline based on which new policies can be inferred. This observation is applicable both for the case of augmenting an existing policy database by incrementally adding new rules as well as creating a new policy for an organization that has a similar structure as that of another organization where an ABAC policy is already deployed. In order to achieve this goal of aiding the policy administration process, we propose the \textit{ABAC Policy Inference Problem} (ABAC-PIP). In this section, we present the formal definition of our proposed problem. 

ABAC-PIP gathers information from an existing policy and then creates new rules based on that learning. This problem variant takes a deployed ABAC policy, a set of subjects, a set of objects, a set of subject attributes and their corresponding values, a set of object attributes and their corresponding values and a set of access requests as inputs and produces as output, for each access request, the set of permissions or operations required to carry out the designated task in case the access request is to be granted or the decision to deny the request if it is an unauthorized one. The formal problem definition is presented below.

\begin{dfn}  \label{abac_pip}

\textbf{ABAC-PIP}\\
Given an ABAC policy $\mathcal{P}$, a set $S$ of subjects, a set $O$ of objects, a set $SA$ of subject attributes, a set $OA$ of object attributes, a function $F_{sub}$ defining the value assignment of the subject attributes of each subject, a function $F_{obj}$ defining the value assignment of object attributes of each object, and a set $\mathcal{L}$ of access requests as inputs, determine, corresponding to each access request, the set of permissions (or operations) $\mathrm{Pr}$, required to execute the access request if the access request is to be permitted or else output the decision to deny the access request if it is not to be permitted.
\end{dfn}

The following underlying assumptions have been made for ABAC-PIP.
\begin{itemize}
\item The access requests present in the set $\mathcal{L}$ are new access requests and the corresponding decisions cannot be determined by the rules present in the already deployed ABAC policy $\mathcal{P}$. 
\item The access requests contained in $\mathcal{L}$ are derived from access logs of an organization.
\item The access requests of $\mathcal{L}$ include both positive and negative access requests. Positive access requests are the ones which are to be granted and negative access requests are those which are to be denied.
\item The new access requests are generated when an organization having a deployed ABAC model undergoes some changes. Such changes necessitate the creation of additional rules corresponding to the new access requests that will take place. An example of such a change can the opening of a new department or the introduction of a new job designation.
\item The new access requests can also be generated when an organization wishes to migrate to ABAC. This organization is similar in structure and workflow to another organization where ABAC has already been deployed.
\item For the new access requests, the relevant information regarding the value assignment for the subject and object attributes are available.
\end{itemize}

In the following section, we discuss the applicability of ABAC-PIP to two policy administration scenarios and present the solution strategy that assists the corresponding policy administration tasks.

\section{Proposed Methodology} \label{framework}

We propose a machine learning based methodology for solving ABAC-PIP. Our approach is capable of helping system administrators in creating ABAC policies and in the process, can reduce the overhead associated with policy creation. We name our proposed framework as \textit{\textbf{\underline{P}}olicy \textbf{\underline{A}}dministration \textbf{\underline{M}}ethodology using \textbf{\underline{M}}achin\textbf{\underline{e}} \textbf{\underline{L}}e\textbf{\underline{a}}rning} (PAMMELA). In this section, we first present a detailed overview of how PAMMELA solves ABAC-PIP and then elaborate on the scenarios where PAMMELA is applicable. We also present two techniques that we have designed to improve the overall performance of PAMMELA.

\subsection{PAMMELA} \label{pammela}

ABAC-PIP has a direct correlation with the process of inferring new ABAC rules. We propose a methodology named as PAMMELA for solving ABAC-PIP. PAMMELA is a supervised learning based solution strategy that works in two phases. In the first phase, a machine learning classifier is trained using a set of labeled data. This labeled data is in the form of an ABAC policy consisting of several rules. Each rule defines the combination of the subject attribute-value pairs and the object attribute-value pairs for which a subject will be allowed to access and perform certain operations or acquire some permissions for an object. We refer to such rules as \textit{positive rules}. 
The attributes are treated as features by the classifier. The positive rules help the classifier in learning under which conditions, an access request is to be granted and what are the corresponding permissions associated with the access. 

In addition to this, it is also essential for the classifier to learn when an access request is to be denied. Such scenarios are covered by the \textit{negative rules}. A negative rule specifies the attribute-value pairings for which an access request is not permissible. For negative rules, all those attribute-value pairs are considered which lead to unauthorized accesses, not just the ones which explicitly deny accesses. Such negative rules can be derived from the set of positive rules. If $\mathcal{U}$ denotes the set of all possible attribute-value pairings (both subject and object) and $\mathcal{PR}$ denotes the set of positive rules, then the set of negative rules $\mathcal{NR}$ = \{$\mathcal{U}$\} $\setminus$ \{$\mathcal{PR}$\}. Though it is straightforward to derive the elements of $\mathcal{NR}$, the task, however, is not trivial. The reason behind this is that, generally in any organization, the set of rules disallowing accesses is much larger in size than the set of rules allowing accesses. In PAMMELA, the machine learning classifier is also trained using the negative rules. In this case, the classifier learns to output denial as a decision.

For supplying the training data, it needs to be encoded in a format that is understandable to the machine learning algorithm. This is done by making use of categorical data. Each subject and object attribute value is assigned a numerical value. For eg., If we have a subject attribute value \textit{Designation} having values \textit{Assistant Professor}, \textit{Associate Professor} and \textit{Professor}, then \textit{Assistant Professor} can be assigned the value 1, \textit{Associate Professor} can be assigned the value 2 and \textit{Professor} can be assigned the value 3. If a particular subject attribute is not applicable for a certain user, then a special value $NA$ (for Not Applicable) is assigned as the attribute value for that subject. For each attribute value set, the categorical encoding starts increasing monotonically from 1. The reason for using categorical encoding is a reduced input vector size in comparison to the vector size that is obtained for one-hot encoding.

After training, PAMMELA generates new rules in the second phase which is the testing phase for the machine learning classifier. Here, a set of access requests is given as input to the classifier. The combinations of the attribute values present in these access requests are different from those present in the existing policy rules. Such new combinations of values can occur when either the subject attribute value set(s) or the object attribute value set(s) or both are augmented with additional values. 

\begin{figure*}[h]
\caption{Block Diagram depicting the workflow of PAMMELA}
\includegraphics[scale = 0.27, trim = 0mm 240mm 20mm 70mm, clip]{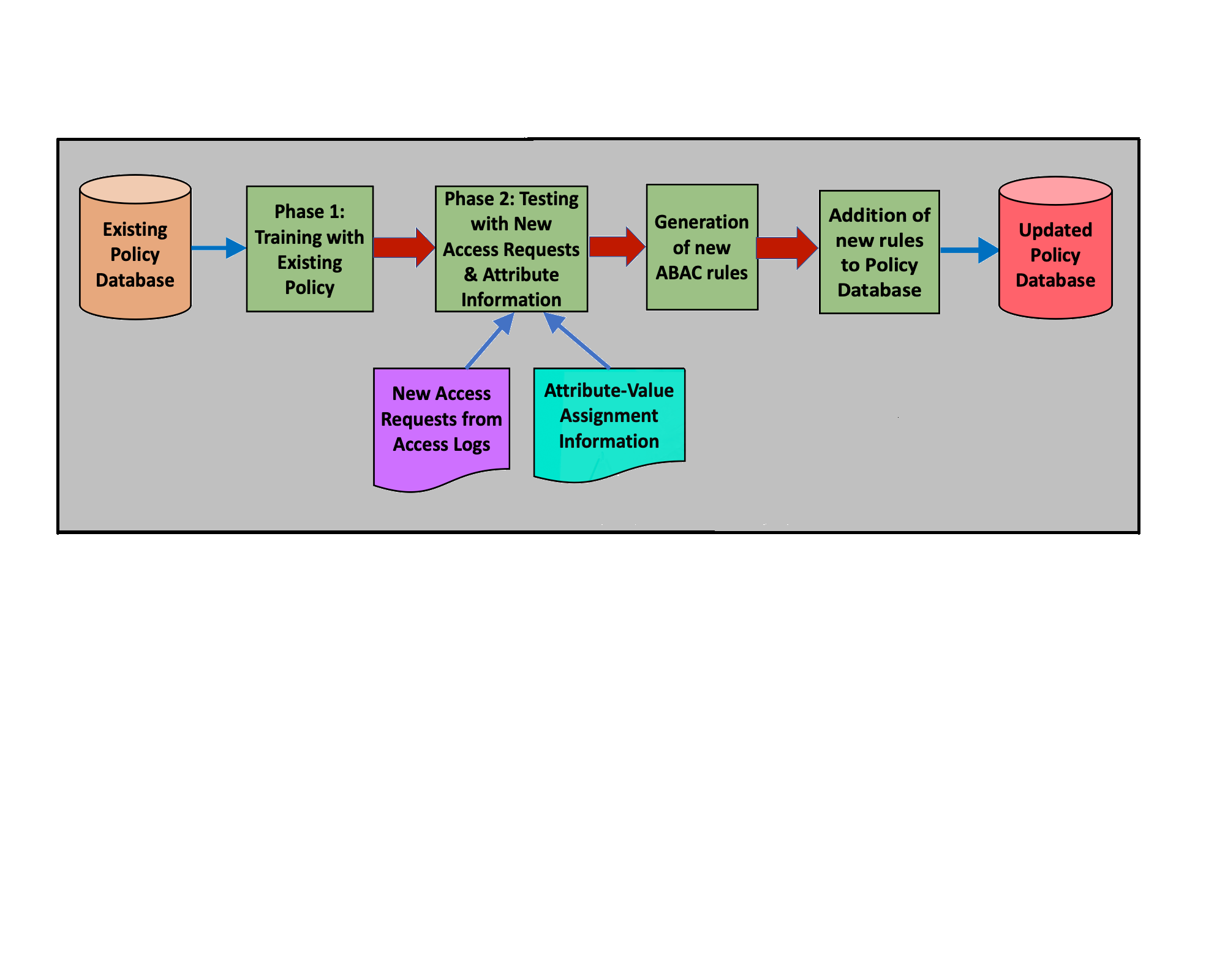}
\label{fig1}
\end{figure*}

Both positive and negative access requests are given as input to PAMMELA in the second phase. Positive access requests are those which are to be granted. Negative access requests are those which are to be denied. The reason for including both positive and negative access requests is that when the already deployed policy cannot account for the organizational changes, then any type of access request needs to be appropriately classified. We assume that for each of the new access requests, the relevant attribute-value pair assignment information are available. Based on the learning achieved in the first phase, the new access requests and the attribute-value information, PAMMELA classifies each access request by determining whether it is to be allowed or disallowed. If the access request is to be disallowed, PAMMELA outputs the decision $NO$. On the other hand, if the access request is to be granted, PAMMELA outputs the set of permissions or operations required to successfully execute the access request in addition to the decision of $YES$. Once the relevant decisions have been derived, the newly generated rules are added to the policy database. The workflow of PAMMELA is depicted in Figure \ref{fig1}.

It needs to be mentioned here that since PAMMELA is trained using an existing set of rules, we do not need to worry about the lack or absence of labeled data. Hence, we can use supervised machine learning techniques in our methodology.

\subsection{Policy Administration via Augmentative Policy Inference} \label{aug}

Undergoing structural and functional changes is not uncommon for any organization. Examples of such changes include but are not limited to the following:
\begin{itemize}
\item An organization can start a new department.
\item An organization can introduce a new job designation.
\item An educational institute can introduce a new degree program.
\item An educational institute can introduce a new undergraduate or graduate level course.
\item A hospital can start treating patients in a new specialization.
\end{itemize}
These types of changes increase the breadth of the workflow of the organization and may not be too frequent. However, whenever they do occur, there arises a need to modify or update the access control policy in order to prevent any form of unauthorized access or leakage of rights. 

Updating the policy by re-executing the policy mining process involves a lot of overhead. Moreover, it increases the overall time required to manage the update process. Hence, it is prudent to design only the new rules and add them to the policy database. The types of changes considered here are not radical enough to completely alter the organizational structure. Hence, the new rules that need to be designed will have some similarity with the existing rules. More specifically, the new rules will differ from the current ones in terms of one or more attribute values. Also, the permission or operation set associated with a new rule will be same as that of an equivalent existing rule. Manually augmenting the policy database is a very challenging task even for a moderate sized organization. 

Our proposed methodology PAMMELA can perform automated augmentation of a deployed policy. We name this application of PAMMELA as \textit{Augmentative Policy Inference}. 
PAMMELA can determine the relevant decisions for both positive and negative accesses, thereby augmenting the policy database by inferring and adding new ABAC rules. Other than providing the categorical encoding for the new attribute values, our proposed approach completely eliminates any human intervention in the policy augmentation process. Of course, the underlying assumption here is that an organization needs to have an already deployed ABAC policy in order to perform policy augmentation using PAMMELA. We assume that the subject attribute value set as well as the object attribute value set are augmented with new values. This kind of attribute value introduction accounts for the different organizational changes. 

We elaborate the observations made above using the following examples. Suppose an $XYZ$ university has two departments - Computer Science and Electronics. The university has deployed the ABAC model for access control. We consider one subject attribute \textit{User-Type} and one object attribute \textit{Resource-Type}. Values of \textit{User-Type} are \textit{Faculty}, \textit{Student} and \textit{Teaching-Assistant}. Values of \textit{Resource-Type} are \textit{Question-Paper}, \textit{Answer-Script}, \textit{Assignment} and \textit{Mark-Sheet}. The university management decides to open a new department, Information Technology. The rules governing the access of the different resources by the users belonging Computer Science and Electronics, will no doubt be applicable for Information Technology. The new rules will have the \textit{Department} attribute value as \textit{Information Technology}. Another example can be that, if the Computer Science department wishes to introduce a new evaluation component, \textit{Quiz}, then the rules for \textit{Quiz} will be similar to those associated with the access of the existing component \textit{Assignment}.

\subsection{Policy Administration via Adaptive Policy Inference} \label{adap}

When an organization intends to migrate to ABAC, a policy needs to be designed ab initio using a policy generation approach. The input to the policy creation process can be an access log consisting of a set of access requests. A considerable amount of administrative overhead associated with this task can be reduced if some already deployed ABAC policy is available which can serve as a guideline or reference point. Henceforth, we shall refer to this existing policy as \textit{reference policy} and the new policy that is to be designed as \textit{target policy}. It is to be noted here that the structure of the organization where the reference policy is deployed is assumed to be similar to the structure of the organization that wishes to implement ABAC. In such a scenario, the rules present in the reference policy can be adapted to generate the target policy. The adaptation process will handle the dissimilarities present between the rules of the reference policy and the target policy in terms of the subject and object attribute values. 

We explain the adaptive policy inference task using an example. We take the example of $XYZ$ university mentioned in Sub-section \ref{aug} having two departments, Computer Science and Electronics. The subject as well as object attributes and their corresponding values are the same as those mentioned in the previous sub-section. Suppose, another educational institute, $PQR$ wishes to adopt the ABAC model. $PQR$ has two departments, Mechanical Engineering and Civil Engineering. $PQR$ has the same subject and object attributes as those of $XYZ$. The subject attribute values of $PQR$ are also same as those of $XYZ$. For the object attribute values, $PQR$ has an additional value \textit{Presentation} apart from those present for $XYZ$. The evaluation component \textit{Presentation} requires the same type of accesses as that of \textit{Assignment}. The rules that need to be implemented for $PQR$ have the same structure as those of $XYZ$ and hence, can be inferred by taking the rules of $XYZ$ as reference. In this way, the reference policy of $XYZ$ can be appropriately adapted to create the target policy for $PQR$.

The above mentioned scenario relates to ABAC-PIP and therefore, can be solved using our proposed methodology. The reference policy will serve as an input to the machine learning classifier in the training phase. In order to generate the target policy, the access requests present in the access logs of the organization that wishes to implement ABAC will be supplied to the classifier in the testing phase. Here, we again assume that both positive and negative requests are given to the classifier. Also, the attribute-value assignment information is available in the access logs or in some other data repository. Based on the learning accomplished during training, the classifier will output a "NO" or a "YES". For a grant decision, the set of permissions associated with the access will also be generated. Finally, the output of the second phase combined with the attribute-value assignment information will create the target ABAC policy. One assumption associated with the adaptive policy inference task using PAMMELA is that the set of permissions or operations associated with each rule of the target policy is the same as that of a corresponding rule of the reference policy.

Several recent works like \cite{saptarshi_adaptation}, \cite{saptarshi_hierarchical}, \cite{polisma} and \cite{flap} have addressed the problem of policy adaptation. The approaches proposed in \cite{saptarshi_adaptation} and \cite{saptarshi_hierarchical} do not make use of machine learning. The techniques in \cite{polisma} and \cite{flap} use machine learning along with a number of heuristic methods. To the best of our knowledge, our proposed methodology PAMMELA is the first end-to-end machine learning based strategy for policy adaptation.

\subsection{Discussion} \label{discuss}

PAMMELA offers the advantage of completely automating the tasks of augmentative as well as adaptive policy inference. Needless to say that if these tasks are carried out manually, it requires a huge amount of effort to painstakingly scan through the policy database to determine which existing rules are most similar to the rules that are to be created and generate the new rules. If the tasks are carried out manually, the entire process becomes very time consuming quite prone to errors. The tasks can be accomplished using heuristic methods but at the cost of designing several functions covering different scenarios and running several experiments to determine the best possible strategy. If each strategy executes for a considerable amount of time, then selection of the most suitable approach will involve a lot of time overhead even before the actual augmentation or adaptation process. PAMMELA eliminates all the above mentioned issues by providing an end-to-end automated solution. This is substantiated in the experimental results presented later.

\subsection{Learning Enhancement Techniques}  \label{let}

In this sub-section, we present two techniques that we have designed to enhance the learning capability of PAMMELA for better performance. The first technique aims at extracting a set of features depending upon the relationships existing among the subject and object attributes. We name this technique as \textit{Attribute Relation based Feature Extraction} (ARFE). The second one deals with the creation of a categorical encoding that is to be fed to the machine learning classifier based on similar types of subject and object attribute values. We name the second approach as \textit{Attribute Value Clustering} (AVC). Details of the two techniques are presented as follows.

\subsubsection{Attribute Relation based Feature Extraction} \label{arfe}

Attribute Relation based Feature Extraction (ARFE) tries to capture the relationships that exist among subject and object attributes. An example of such a relation is - \textit{A student registered to the Database Systems course can access the assignment of the same course}. Here, the attribute \textit{Course} is associated with both the subject \textit{Student} as well as the object \textit{Assignment}. This implies that if an attribute is present both as a subject and an object attribute and have the same value, then most likely, the subject is eligible to access the object. We use the term most likely because the final access decision can be dependent upon other attribute values or similar other such attribute relations. 

In order to capture this type of relationship between pairs of common subject and object attributes, we introduce additional features. Suppose an attribute $Attr$ is present in a rule or an access request as a subject as well as an object attribute. We denote $Attr$ as a subject attribute by $S\_Attr$ and $Attr$ as an object attribute by $O\_Attr$. We introduce a new feature $F_{Attr}$ which assumes different values based upon the attribute values of $S\_Attr$ and $O\_Attr$. The number of new features introduced is equal to the number of such common subject and object attributes.

\subsubsection{Attribute Value Clustering} \label{avc}

In Attribute Value Clustering (AVC), we group the different values of a particular attribute based on their similarity. Suppose we have an attribute \textit{Resource-Type} which has the following values - \textit{Quiz}, \textit{Assignment}, \textit{Question-Paper}, \textit{Answer-Script}, \textit{Department-Office-Record} and \textit{Department-Budget}. It can be easily concluded that \textit{Quiz} and \textit{Assignment}, \textit{Question-Paper} and \textit{Answer-Script} and \textit{Department-Office-Record} and \textit{Department-Budget} are similar types of resources. Such similarity implies same types of accesses. Consequently, the rules governing the access to these similar types of resources will follow the same structure and format. This observation holds true for any type of attribute.

AVC attempts to capture the grouping of attribute values of any attribute depending upon the functional alikeness among the values. This clustering of similar attribute values is captured in the categorical encoding used in PAMMELA. We feel that the use of AVC in conjunction with PAMMELA will improve the overall performance of deriving new rules.

\section{Experimental Results} \label{exp}

In this section, we evaluate the performance of our proposed approach. First, we present a description of the datasets that we have used for experimental evaluation. We then discuss about the metrics used for evaluating the performance of PAMMELA. Finally, we report the experimental findings.

\subsection{Dataset Description} \label{dataset}

For the purpose of experimentation, we have manually created three datasets. Each dataset consists of two parts:
\begin{itemize}
\item An ABAC policy consisting of a number of rules 
\item A set of access requests along with the attribute-value assignment information
\end{itemize}

The first part of the dataset, i.e., the ABAC policy is used in the training phase of PAMMELA. This corresponds to the existing or the reference policy that we have mentioned earlier. The set of access requests and the attribute-value assignment information are used in the testing phase. 

We have not used any synthetic dataset generator or simulator for creating the datasets. The datasets have been created keeping real-world scenarios in mind. We have generated the set of subjects, objects, subject attributes, object attributes and their corresponding values in such a way that they mimic real-life organizations. The rules as well as the access requests of each dataset are similar to actual organizational accesses. For each of the datasets, we have considered only one permission corresponding to the ability of a subject to access a given resource or the denial of access. However, PAMMELA is capable of handling multi-permission scenarios. 

The different aspects of each dataset are described below:\\
\textbf{University Dataset 1:} \\
We have designed a dataset to mimic the workflow of an educational institute and have named it as \textit{University Dataset 1}. It contains the following subject attributes - \textit{Designation}, \textit{Department}, \textit{Degree} and \textit{Year}. The object attributes are \textit{Resource-Type}, \textit{Department}, \textit{Degree} and \textit{Year}. Table \ref{1} shows the details regarding the number of values associated with each attribute of the dataset. The training data contains 53 rules. The test data contains 1010 access requests out of which 598 are positive requests and 412 are negative requests. In the test data, the subject attributes for which new values have been added are \textit{Designation}, \textit{Department}, and \textit{Degree} and the object attributes for which new values have been introduced are \textit{Resource-Type}, \textit{Department} and \textit{Degree}. \\
\begin{table}\center
\caption{Attribute-Value Count for University Dataset 1}
\label{1}
\begin{tabular}{l|c|c}
\textbf{Attribute-Name} & \textbf{Attribute-Type} & \textbf{No. of Values} \\
\hline
\hline
Designation & Subject & 3 \\
Department & Subject & 4 \\
Degree & Subject & 2 \\
Year & Subject & 4 \\
Resource-Type & Object & 7 \\
Department & Object & 4 \\
Degree & Object & 2 \\
Year & Object & 4 \\
\hline
\end{tabular}
\end{table}
\textbf{University Dataset 2:} \\
Another dataset has been designed keeping in mind the structure of an educational institute and has been named as \textit{University Dataset 2}. However, there exists several differences between this dataset and the previous one. \textit{University Dataset 2 } is much more detailed than \textit{University Dataset 1} both in terms of the attributes, their corresponding values as well as the number of rules present in the training data. The subject attributes considered in this dataset are \textit{Designation}, \textit{Post}, \textit{Course}, \textit{Department}, \textit{Degree} and \textit{Year}. The object attributes are \textit{Resource-Type}, \textit{Course}, \textit{Department}, \textit{Degree} and \textit{Year}. Table \ref{2} shows the details regarding the number of values associated with each attribute of the dataset. The training data contains 1,56,775 rules and the test data contains 483 access requests out of which 308 are positive access requests and 175 are negative access requests. In the test data, the subject attributes for which new values have been added are \textit{Designation}, \textit{Post},  \textit{Course} and \textit{Department} and the object attributes for which new values have been introduced are \textit{Resource-Type}, \textit{Course} and \textit{Department}.\\
\begin{table}\center
\caption{Attribute-Value Count for University Dataset 2}
\label{2}
\begin{tabular}{l|c|c}
\textbf{Attribute-Name} & \textbf{Attribute-Type} & \textbf{No. of Values} \\
\hline
\hline
Designation & Subject & 5 \\
Post & Subject & 2 \\
Department & Subject & 5 \\
Course & Subject & 120 \\
Degree & Subject & 2 \\
Year & Subject & 4 \\
Resource-Type & Object & 8 \\
Department & Object & 5 \\
Course & Object & 120 \\
Degree & Object & 2 \\
Year & Object & 4 \\
\hline
\end{tabular}
\end{table}
\textbf{Company Dataset:}\\
This dataset has been created keeping in mind the structure and workflow of a software company. The dataset may not be as comprehensive and extensive as an actual company. Nonetheless, it incorporates features and access rules similar to a real-life organization. The dataset contains three subject attributes, \textit{Designation}, \textit{Project-Name} and \textit{Department} and three object attributes, \textit{Resource-Type}, \textit{Project-Name} and \textit{Department}. The details regarding the number of values associated with each attribute are shown in Table \ref{3}. There are 384 rules present in the training data. 291 access requests are present in the test data out of which 93 are positive access requests and 198 are negative ones. For subject attributes \textit{Designation} and \textit{Project-Name} and object attributes \textit{Resource-Type} and \textit{Project-Name}, new values have been introduced in the test data.\\
\begin{table}\center
\caption{Attribute-Value Count for Company Dataset}
\label{3}
\begin{tabular}{l|c|c}
\textbf{Attribute-Name} & \textbf{Attribute-Type} & \textbf{No. of Values} \\
\hline
\hline
Designation & Subject & 12 \\
Project-Name & Subject & 2 \\
Department & Subject & 5 \\
Resource-Type & Object & 8 \\
Project-Name & Object & 2 \\
Department & Object & 4 \\
\hline
\end{tabular}
\end{table}

It is to be noted here that if a certain attribute is not applicable for a particular subject or object, then the value \textit{Not Applicable} is assigned for that attribute. The number of rules and access requests present respectively in the training and test data have no correlation with the dataset type. These numbers have been chosen purely as a design choice. We have created the datasets in order to cover three scenarios - (i) the number of rules present in training data is less than the number of access requests present in test data, (ii) the number of rules present in training data is higher than the number of access requests present in the test data, and (iii) the number of rules present in training data is comparable to the number of access requests of the test data. Also, in the test data of each dataset, one new attribute value has been introduced in each access request. 

\subsection{Evaluation Metrics} \label{metric}

For evaluating the performance of PAMMELA, we use the following metrics - Accuracy, Precision, Recall and F1-score. Before discussing about these metrics, we introduce few terminologies that are used in the computation of the metrics. The definitions of the terminologies as well as the metrics exist in the domain of machine learning. However, here we present the definitions for the sake of completeness and also making them meaningful in the context of the current work. We use $YES$ and $NO$ to respectively denote the grant decision and the deny decision given as output by PAMMELA. The terminologies and the metrics are defined as follows:
\begin{itemize}
\item \textbf{True Positive Accesses (TPA):} These are the positive access requests corresponding to which PAMMELA gives the output $YES$. These are instances of correct classifications.
\item \textbf{True Negative Accesses (TNA):} These are the negative access requests corresponding to which PAMMELA gives the output $NO$. These are instances of correct classifications.
\item \textbf{False Positive Accesses (FPA):} These are the negative access requests corresponding to which PAMMELA gives the output $YES$. These are instances of misclassifications resulting in security breach.
\item \textbf{False Negative Accesses (FNA):} These are the positive access requests corresponding to which PAMMELA gives the output $NO$. These are instances of misclassifications resulting in an over-restrictive system.
\item \textbf{Accuracy:} It is the ratio of the correctly classified access requests to the total number of access requests. It denotes the capability of the classifier to make correct decisions. Mathematically, 
\begin{equation}
Accuracy = \frac{TPA + TNA}{TPA + FPA + TNA + FNA}
\end{equation}
\item \textbf{Precision:} It is the ratio of the correctly classified positive access requests to the total number of access requests for which the output is $YES$. Precision is inversely proportional to the degree of security breach occurring in the system. Mathematically,
\begin{equation}
Precision = \frac{TPA}{TPA + FPA}
\end{equation}
\item \textbf{Recall:} It is the ratio of the correctly classified positive access requests to the total number of positive access requests. Recall is inversely proportional to the degree of over-restrictiveness of the system. Mathematically,
\begin{equation}
Recall = \frac{TPA}{TPA + FNA}
\end{equation}
\item \textbf{F1-score:} It is calculated as the weighted average of precision and recall. F1-score balances the relative trade-off between precision and recall. It is calculated as
\begin{equation}
F1-score = \frac{2 * Precision * Recall}{Precision + Recall}
\end{equation}

\end{itemize}

\subsection{Results} \label{res}

In this section, we present the experimental results. We have used the following machine learning classifiers to evaluate the performance of PAMMELA - Artificial Neural Network (ANN) \cite{Murphy}, Decision Tree (DT) \cite{DecisionTrees}, Random Forest (RF) \cite{RandomForest}, Extra Trees (ET) \cite{ExtraTrees}, GradientBoosting (GB) \cite{scikit-learn} and XGBoost (XGB) \cite{scikit-learn}. We report the results of the following four approaches.
\begin{itemize}
\item \textbf{PAMMELA without any enhancement} referred to as PAMMELA-Naive
\item \textbf{PAMMELA with Attribute Relation based Feature Extraction} referred to as PAMMELA-ARFE
\item \textbf{PAMMELA with Attribute Value Clustering} referred to as PAMMELA-AVC
\item \textbf{PAMMELA with Attribute Relation based Feature Extraction and Attribute Value Clustering} referred to as PAMMELA-ARFE+AVC
\end{itemize}
The implementation has been done using Python 3.8 and the experiments were conducted on a MacBook Pro laptop having 2.3 GHz, 8 cores, intel core i9 processor, 16 GB RAM and macOS 11.4 as operating system. We report the performance of the four approaches in terms of the metrics Accuracy (Acc), Precision (Pr), Recall (Rec) and F1-score (F1-s). Tables  \ref{university1} - \ref{company} present the results of the three datasets. In each table, the first column (Clfr) denotes the name of the machine learning classifier used, each represented using the abbreviations mentioned earlier. Also, the names of the four approaches and the metrics are denoted using the compressed forms mentioned above.

\begin{table*}[]\scriptsize
\caption{Experimental Results for University Dataset 1}
\label{university1}
\begin{tabular}{|l|l|l|l|l|l|l|l|l|l|l|l|l|l|l|l|l|}
\hline
\textbf{Clfr} & \multicolumn{4}{c|}{\textbf{PAMMELA-Naive}} & \multicolumn{4}{c|}{\textbf{PAMMELA-ARFE}} & \multicolumn{4}{c|}{\textbf{PAMMELA-AVC}} & \multicolumn{4}{c|}{\textbf{PAMMELA-ARFE+AVC}} \\
\cline{2-17}
           & \textit{Acc} & \textit{Pr} & \textit{Rec} & \textit{F1-s} & \textit{Acc} & \textit{Pr} & \textit{Rec} & \textit{F1-s} & \textit{Acc} & \textit{Pr} & \textit{Rec} & \textit{F1-s} & \textit{Acc} & \textit{Pr} & \textit{Rec} & \textit{F1-s}  \\ \hline \hline
           
ANN & 0.786    & 0.934     & 0.687  & 0.792   & 0.905    & 0.925     & 0.913  & 0.919   & 0.853    & 0.975     & 0.773  & 0.862   & 0.998    & 0.997     & 1.000  & 0.998  \\

DT           & 0.864    & 0.932     & 0.831  & 0.879   & 0.982    & 0.977     & 0.993  & 0.985   & 0.978    & 0.972     & 0.992  & 0.982   & 0.994    & 0.997     & 0.993  & 0.995    \\

RF          & 0.876    & 0.944     & 0.841  & 0.889   & 0.982    & 0.977     & 0.993  & 0.985   & 0.968    & 0.965     & 0.982  & 0.973   & 0.994    & 0.997     & 0.993  & 0.995  \\ 

ET           & 0.882    & 0.967     & 0.829  & 0.893   & 0.982    & 0.977     & 0.993  & 0.985   & 0.972    & 0.967     & 0.987  & 0.977   & 0.994    & 0.997     & 0.993  & 0.995  \\ 

GB       & 0.881    & 0.935     & 0.860  & 0.895   & 0.982    & 0.977     & 0.993  & 0.985   & 0.943    & 0.972     & 0.930  & 0.950   & 0.994    & 0.997     & 0.993  & 0.995  \\ 

XGB                & 0.864    & 0.932     & 0.831  & 0.879   & 0.982    & 0.977     & 0.993  & 0.985   & 0.976    & 0.967     & 0.993  & 0.980   & 0.994    & 0.997     & 0.993  & 0.995  \\ \hline
\end{tabular}
\end{table*}

\begin{table*}[]\scriptsize
\caption{Experimental Results for University Dataset 2}
\label{university2}
\begin{tabular}{|l|l|l|l|l|l|l|l|l|l|l|l|l|l|l|l|l|}
\hline
\textbf{Clfr} & \multicolumn{4}{c|}{\textbf{PAMMELA-Naive}} & \multicolumn{4}{c|}{\textbf{PAMMELA-ARFE}} & \multicolumn{4}{c|}{\textbf{PAMMELA-AVC}} & \multicolumn{4}{c|}{\textbf{PAMMELA-ARFE+AVC}} \\
\cline{2-17}
           & \textit{Acc} & \textit{Pr} & \textit{Rec} & \textit{F1-s} & \textit{Acc} & \textit{Pr} & \textit{Rec} & \textit{F1-s} & \textit{Acc} & \textit{Pr} & \textit{Rec} & \textit{F1-s} & \textit{Acc} & \textit{Pr} & \textit{Rec} & \textit{F1-s}  \\ \hline \hline
           
ANN & 0.590    & 0.958     & 0.373  & 0.537   & 0.907    & 1.000     & 0.854  & 0.921   & 0.642    & 0.993     & 0.442  & 0.611   & 0.954    & 1.000     & 0.929  & 0.963   \\ 

DT          & 0.627    & 0.971     & 0.429  & 0.595   & 0.915    & 1.000     & 0.867  & 0.929   & 0.648    & 0.973     & 0.461  & 0.626   & 0.957    & 1.000     & 0.932  & 0.965   \\

RF           & 0.555    & 0.943     & 0.321  & 0.479   & 0.915    & 1.000     & 0.867  & 0.929   & 0.576    & 0.964     & 0.347  & 0.511   & 0.957    & 1.000     & 0.932  & 0.965   \\

ET             & 0.551    & 0.925     & 0.321  & 0.477   & 0.915    & 1.000     & 0.867  & 0.929   & 0.596    & 0.945     & 0.390  & 0.552   & 0.957    & 1.000     & 0.932  & 0.965     \\

GB       & 0.331    & 0.174     & 0.013  & 0.024   & 0.393    & 0.551     & 0.263  & 0.356   & 0.362    & 0.500     & 0.062  & 0.110   & 0.973    & 1.000     & 0.958  & 0.978   \\ 

XGB                & 0.555    & 0.979     & 0.308  & 0.469   & 0.913    & 1.000     & 0.864  & 0.927   & 0.491    & 0.970     & 0.208  & 0.342   & 0.965    & 1.000     & 0.945  & 0.972   \\ \hline
\end{tabular}
\end{table*}

\begin{table*}[]\scriptsize
\caption{Experimental Results for Company Dataset}
\label{company}
\begin{tabular}{|l|l|l|l|l|l|l|l|l|l|l|l|l|l|l|l|l|}
\hline
\textbf{Clfr} & \multicolumn{4}{c|}{\textbf{PAMMELA-Naive}} & \multicolumn{4}{c|}{\textbf{PAMMELA-ARFE}} & \multicolumn{4}{c|}{\textbf{PAMMELA-AVC}} & \multicolumn{4}{c|}{\textbf{PAMMELA-ARFE+AVC}} \\
\cline{2-17}
           & \textit{Acc} & \textit{Pr} & \textit{Rec} & \textit{F1-s} & \textit{Acc} & \textit{Pr} & \textit{Rec} & \textit{F1-s} & \textit{Acc} & \textit{Pr} & \textit{Rec} & \textit{F1-s} & \textit{Acc} & \textit{Pr} & \textit{Rec} & \textit{F1-s}  \\ \hline \hline

ANN & 0.732    & 0.584     & 0.559  & 0.571   & 0.698    & 0.532     & 0.441  & 0.482   & 0.729    & 0.646     & 0.333  & 0.440   & 0.808    & 0.718     & 0.656  & 0.685     \\
 
DT           & 0.790    & 0.627     & 0.849  & 0.721   & 0.842    & 0.701     & 0.882  & 0.781   & 0.955    & 0.877     & 1.000  & 0.935   & 1.000    & 1.000     & 1.000  & 1.000     \\ 

RF           & 0.856    & 0.726     & 0.882  & 0.796   & 0.835    & 0.689     & 0.882  & 0.774   & 0.955    & 0.877     & 1.000  & 0.935   & 1.000    & 1.000     & 1.000  & 1.000    \\

ET             & 0.859    & 0.728     & 0.892  & 0.802   & 0.838    & 0.698     & 0.871  & 0.775   & 0.955    & 0.877     & 1.000  & 0.935   & 1.000    & 1.000     & 1.000  & 1.000  \\

GB       & 0.797    & 0.673     & 0.710  & 0.691   & 0.838    & 0.725     & 0.796  & 0.759   & 0.845    & 0.800     & 0.688  & 0.740   & 1.000    & 1.000     & 1.000  & 1.000    \\ 

XGB           & 0.753    & 0.624     & 0.570  & 0.596   & 0.863    & 0.762     & 0.828  & 0.794   & 0.863    & 0.853     & 0.688  & 0.762   & 1.000    & 1.000     & 1.000  & 1.000    \\ \hline
\end{tabular}
\end{table*}

The results show that, in general, the performance of PAMMELA-ARFE, PAMMELA-AVC as well as PAMMELA-ARFE+AVC is better than that of PAMMELA-Naive in terms of F1-score. This substantiates the fact that the proposed learning enhancement techniques, either used in isolation or in combination, improve the overall performance of PAMMELA. For the University Datasets , PAMMELA-ARFE outperforms PAMMELA-AVC. The reason behind this is that, in these two datasets, several common subject and object attributes are present and the corresponding relations among the attributes are efficiently captured by the ARFE strategy. However, for the Company Dataset, the performance of PAMMELA-AVC is better than PAMMELA-ARFE for F1-score values. This happens because of the nature of the dataset. The Company Dataset has only two common subject and object attributes. Moreover, several positive rules present in the Company Dataset are such that the values for the common subject and object attributes are not same. Hence, the ARFE strategy cannot much enhance the performance of the approach. However, for this dataset, the AVC technique, in general, improves the performance over the Naive strategy though the improvement is not very substantial. For all the datasets, PAMMELA-ARFE+AVC gives better results than either of PAMMELA-ARFE and PAMMELA-AVC in terms of F1-score. This shows that the two enhancement strategies when used simultaneously give the best possible performance among all the competing approaches.

The above mentioned observations are true for most of the machine learning classifiers. In order to highlight the performance gain provided by the enhancement techniques, we select GradientBoosting (GB). For University Dataset 1, in terms of the F1-score, PAMMELA-AVC, PAMMELA-ARFE and PAMMELA-ARFE+AVC give a performance gain respectively of 5.5\%, 9\% and 10\% for GB. For the University Dataset 2 and for GB, PAMMELA-AVC, PAMMELA-ARFE and PAMMELA-ARFE+AVC improve the value of F1-score over the Naive approach respectively by 8.6\%, 33.2\% and 95.4\%. This substantial amount of performance enhancement of PAMMELA-ARFE+AVC can be attributed to the fact that the University Dataset 2 is the largest of the three datasets both in terms of the number of attributes as well as the number of rules and these aspects of the dataset are effectively handled by the two combined learning enhancement strategies. For the Company Dataset, in terms of F1-score, PAMMELA-ARFE, PAMMELA-AVC and PAMMELA-ARFE+AVC improve the performance over PAMMELA-Naive respectively by 6.8\%, 4.9\% and 30.9\% for GB.

Based on the above observations, we provide the following insights for system administrators for managing ABAC policies using PAMMELA.
\begin{itemize}
\item A system administrator can experiment with a number of classifiers for PAMMELA before selecting the best possible option.
\item If a dataset contains several common subject and object attributes and if the administrator wants to use only one enhancement strategy, then it is better to use ARFE rather than AVC.
\item If a dataset does not contain too many common subject and object attributes and a number of rules where the common attributes have not been assigned the same value, then it is better to use AVC rather than ARFE.
\item If a system administrator wants to get the best possible result, then ARFE+AVC should be chosen. However, some amount of effort is required for the feature extraction and the value clustering.
\end{itemize}

The training and the testing time for PAMMELA is quite low. We observed that for ANN, PAMMELA-AVC took 86 seconds to complete the training on the University Dataset 2. This is the highest recorded training time for our experiments. For any other combination of strategy and classifier, PAMMELA gave much lower training execution times on all the datasets. The maximum testing time for PAMMELA that we observed was less than 1 second. Of course, the testing time is dependent upon the number of access requests considered. Hence, experimenting with multiple classifiers will not be too time consuming. Moreover, once training has been done, PAMMELA can keep on generating new rules without any further training. However, if the policy used for training needs to be changed altogether (may happen if the organizational structure or workflow undergoes some radical modifications), then the classifier needs to be re-trained.

\section{Conclusion and Future Scope} \label{concl}

In this paper, we have proposed the ABAC Policy Inference Problem (ABAC-PIP). ABAC-PIP aims to derive a new set of attribute based rules from an existing policy. We have proposed an end-to-end supervised learning based, automated methodology, PAMMELA for solving ABAC-PIP. System administrators can use PAMMELA for augmentative as well as adaptive policy inference when an organization undergoes some changes or an organization migrates to ABAC by adapting the policy of a similar organization. We have also designed two techniques to enhance the learning capability of PAMMELA. Our experimental results show that PAMMELA can be effectively used for the application scenarios mentioned above. Moreover, the learning enhancement techniques indeed improve the performance of PAMMELA.

In future, we intend to apply deep learning, reinforcement learning and incremental learning for inferring ABAC policies. Another direction of future research can be attempting to adapt the ABAC policies of multiple organizations for a single  target organization. This will be a challenging task since it will necessitate resolving the conflicts among the rules of different organizations in order to generate the target policy.

\bibliographystyle{plain}
\bibliography{asiaccs.bib}

\begin{thebibliography}{10}

\bibitem{constraint_nlp}
M.~Alohaly, H.~Takabi, and E.~Blanco.
\newblock Towards an automated extraction of abac constraints from natural
  language policies.
\newblock In {\em Proc. of ICT Systems Security and Privacy Protection}, pages
  105--119, 2019.

\bibitem{policy_recon}
G.~Batra, V.~Atluri, J.~Vaidya, and S.~Sural.
\newblock In-memory policy indexing for policy retrieval points in
  attribute-based access control.
\newblock In {\em Proc. of Int. Conf. on Information Systems Security}, pages
  99--120, 2021.

\bibitem{incr_abac}
G.~Batra, V.~Atluri, J.~Vaidya, and S.~Sural.
\newblock Incremental maintenance of abac policies.
\newblock In {\em Proc. of the 11th ACM Conf. on Data and Application Security
  and Privacy}, pages 185--196, 2021.

\bibitem{RandomForest}
L.~Breiman.
\newblock Random forests.
\newblock {\em Machine Learning}, 45(1):5--32, 2001.

\bibitem{cotrini}
C.~Cotrini, T.~Weghorn, and D.~Basin.
\newblock Mining abac rules from sparse logs.
\newblock In {\em Proc. of 2018 IEEE European Symposium on Security and
  Privacy}, pages 31--46, 2018.

\bibitem{saptarshi_adaptation}
S.~Das, S.~Sural, J.~Vaidya, and V.~Atluri.
\newblock Policy adaptation in attribute-based access control for
  inter-organizational collaboration.
\newblock In {\em Proc. of IEEE 3rd Int. Conf. on Collaboration and Internet
  Computing}, pages 136--145, 2017.

\bibitem{hype}
S.~{Das}, S.~{Sural}, J.~{Vaidya}, and V.~{Atluri}.
\newblock Hype: A hybrid approach toward policy engineering in attribute-based
  access control.
\newblock {\em IEEE Letters of the Computer Society}, pages 25--29, 2018.

\bibitem{saptarshi_hierarchical}
S.~Das, S.~Sural, J.~Vaidya, and V.~Atluri.
\newblock Policy adaptation in hierarchical attribute-based access control
  systems.
\newblock {\em ACM Trans. on Internet Technology}, 19(3), August 2019.

\bibitem{constrainedmining}
M.~Gautam, S.~Jha, S.~Sural, J.~Vaidya, and V.~Atluri.
\newblock Poster: Constrained policy mining in attribute based access control.
\newblock In {\em ACM Symposium on Access Control Models and Technologies},
  pages 121--123, 2017.

\bibitem{ExtraTrees}
P.~Geurts, D.~Ernst, and L.~Wehenkel.
\newblock Extremely randomized trees.
\newblock {\em Machine Learning}, 63:3--42, 2006.

\bibitem{dac}
M.~A. Harrison, W.~L. Ruzzo, and J.~D. Ullman.
\newblock Protection in operating systems.
\newblock {\em Communications of the ACM}, 19(8):461–471, 1976.

\bibitem{policy_indexing}
D.~Heutelbeck, M.~L. Baur, and M.~Kluba.
\newblock In-memory policy indexing for policy retrieval points in
  attribute-based access control.
\newblock In {\em Proc. of 26th ACM Symposium on Access Control Models and
  Technologies}, pages 59--70, 2021.

\bibitem{abac_nist}
V.~C. Hu, D.~Ferraiolo, D.~R. Kuhn, A.~Schnitzer, K.~Sandlin, R.~Miller, and
  K.~Scarfone.
\newblock Guide to {A}ttribute-{B}ased {A}ccess {C}ontrol ({ABAC}) definition
  and considerations.
\newblock Technical report, NIST Special Publication, 2014.

\bibitem{polisma}
A.~A. Jabal, E.~Bertino, J.~Lobo, M.~Law, A~Russo, S.~Calo, and D.~Verma.
\newblock Polisma - a framework for learning attribute-based access control
  policies.
\newblock In {\em Proc. of European Symposium on Research in Computer
  Security}, volume 12308, pages 523--544, 2020.

\bibitem{flap}
A.~A. Jabal, E.~Bertino, J.~Lobo, D.~Verma, S.~Calo, and A.~Russo.
\newblock Flap -- a federated learning framework for attribute-based access
  control policies, 2020.

\bibitem{abac_reinforcement}
L.~Karimi, M.~Abdelhakim, and J.~Joshi.
\newblock Adaptive abac policy learning: A reinforcement learning approach,
  2021.

\bibitem{joshi_log}
L.~Karimi, M.~Aldairi, J.~Joshi, and M.~Abdelhakim.
\newblock An automatic attribute based access control policy extraction from
  access logs.
\newblock {\em IEEE Trans. on Dependable and Secure Computing}, 2021.

\bibitem{abac_policy_engg}
L.~Krautsevich, A.~Lazouski, F.~Martinelli, and A.~Yautsiukhin.
\newblock Towards {A}ttribute-{B}ased {A}ccess {C}ontrol policy engineering
  using risk.
\newblock In {\em Proc. of 1st Int. Workshop on Risk Assessment and Risk-Driven
  Testing}, pages 80--90, November 2013.

\bibitem{policy_review}
S.~Lawal and R.~Krishnan.
\newblock Enabling flexible administration in abac through policy review: A
  policy machine case study.
\newblock In {\em Proc. of 7th IEEE Int. Conf. on Big Data Security on Cloud,
  IEEE Int. Conf. on High Performance and Smart Computing, and IEEE Int. Conf.
  on Intelligent Data and Security}, pages 69--74, 2021.

\bibitem{logdeep}
D.~C. Mocanu, F.~Turkmen, and A.Liotta.
\newblock Towards abac policy mining from logs with deep learning.
\newblock {\em Intelligent Systems}, pages 124--128, 2015.

\bibitem{Murphy}
K.~P. Murphy.
\newblock {\em Machine Learning: A Probabilistic Perspective}.
\newblock The MIT Press, 2012.

\bibitem{topdown}
M.~Narouei, H.~Khanpour, H.~Takabi, N.~Parde, and R.~Nielsen.
\newblock Towards a top-down policy engineering framework for attribute-based
  access control.
\newblock {\em ACM Symposium on Access Control Models and Technologies}, pages
  103--114, 2017.

\bibitem{scikit-learn}
F.~Pedregosa, G.~Varoquaux, A.~Gramfort, V.~Michel, B.~Thirion, O.~Grisel,
  M.~Blondel, P.~Prettenhofer, R.~Weiss, V.~Dubourg, J.~Vanderplas, A.~Passos,
  D.~Cournapeau, M.~Brucher, M.~Perrot, and E.~Duchesnay.
\newblock Scikit-learn: Machine learning in {P}ython.
\newblock {\em Journal of Machine Learning Research}, 12:2825--2830, 2011.

\bibitem{DecisionTrees}
J.R. Quinlan.
\newblock Simplifying decision trees.
\newblock {\em Int. Journal of Man-Machine Studies}, 27(3):221--234, 1987.

\bibitem{abac_least_privilege}
M.~W. Sanders and C.~Yue.
\newblock Mining least privilege attribute based access control policies.
\newblock In {\em Proc. of the 35th Annual Computer Security Applications
  Conf.}, pages 404--416, 2019.

\bibitem{rbac}
R.~S. Sandhu, E.~J. Coyne, H.~L. Feinstein, and C.~E. Youman.
\newblock {R}ole-{B}ased {A}ccess {C}ontrol {M}odels.
\newblock {\em IEEE Computer}, 29(2):38--47, 1996.

\bibitem{mac}
R.S. Sandhu.
\newblock Lattice-based access control models.
\newblock {\em Computer}, 26(11):9--19, 1993.

\bibitem{efficient_mining}
T.~Talukdar, G.~Batra, J.~Vaidya, V.~Atluri, and S.~Sural.
\newblock Efficient bottom-up mining of attribute based access control
  policies.
\newblock {\em IEEE Int. Conf. on Collaboration and Internet Computing}, pages
  339--348, 2017.

\bibitem{logs}
Z.~Xu and S.~D. Stoller.
\newblock Mining attribute-based access control policies from logs.
\newblock In {\em Proc. of the 28th Annual IFIP WG 11.3 Working Conf. on Data
  and Applications Security and Privacy}, pages 276--291, 2014.

\bibitem{mining}
Z.~Xu and S.D. Stoller.
\newblock Mining attribute-based access control policies.
\newblock {\em IEEE Trans. on Dependable and Secure Computing}, 12(5):533--545,
  2015.

\end{thebibliography}



\end{document}